# Evaluating the diversity of scientific discourse on twenty-one multilingual Wikipedias using citation analysis

Michael Taylor (University of Wolverhampton, Digital Science) - correspondent, 61 Home Close, Oxford OX2 8PT, m.r.taylor@wlv.ac.uk; m.taylor@digital-science.com;
Roisi Proven (ex-Altmetric);
Carlos Areia (Digital Science, University of Coventry). 0000-0002-4668-7069

# Abstract

**INTRODUCTION:** Wikipedia is a major source of information, particularly for medical and health content, citing over 4 million scholarly publications. However, the representation of research-based knowledge across different languages on Wikipedia has been under explored. This study analyses the largest database of Wikipedia citations collected to date, examining the uniqueness of content and research representation across languages.

**METHOD:** The study included nearly 3.5 million unique research articles and their Wikipedia mentions from 21 languages. These were categorized into three groups: Group A (publications uniquely cited by a single non-English Wikipedia), Group B (co-cited by English and non-English Wikipedias), and Group C (co-cited by multiple non-English Wikipedias). Descriptive and comparative statistics were conducted by Wikipedia language, group, and discipline.

**RESULTS:** Significant differences were found between twenty non-English languages and English Wikipedia ($p<0.001$). While English Wikipedia is the largest, non-English Wikipedias cite an additional 1.5 million publications.

**CONCLUSION:** English Wikipedia should not be seen as a comprehensive body of information. Non-English Wikipedias cover unique subjects and disciplines, offering a more complete representation of research collectively. The uniqueness of voice in non-English Wikipedias correlates with their size, though other factors may also influence these differences.

# Conflicts of interest

# Introduction

Wikipedia, an internet encyclopaedia, was launched in 2001 (Wikipedia, 2022a) and by 2009, held 3 million articles in English, being maintained by just under 500,000 editors (Arthur, 2009). Wikipedia was recorded as being the seventh most visited website in the world in April 2023 (Top websites ranking, n.d). The first non-English Wikipedias were also launched in 2001, including German, Russian, Chinese, Catalan, Japanese (Wikipedia, 2022b). By February 2022, Wikipedia versions were available in 326 languages. Recently, by article count, the largest six are English, German, French, Dutch and Italian (Epstein, 2022). The English-language Wikipedia currently consists of approximately 6.5M articles in English from a total of 60M articles across all languages (Wikipedia, 2022c). The bulk of contributions to Wikipedia pages are created by a small proportion of editors; although a general phenomenon, this skewness is highest for the English and German Wikipedia (Ortega et al., 2008). Coverage of the non-English language Wikipedias has been criticised for lack of inclusion and diversity, with proposals being made to further address this apparent deficit (Wulczyn et al., 2016).

Scientific articles, and in particular, medical topics, are amongst the most visited Wikipedia pages (Smith, 2020). Nevertheless, Wikipedia contains a disclaimer, which asserts that most of its content is written by non-professional editors (Wikipedia, 2022d). Evidence of possible subjectivity of editors has been identified through a correlation study, which examined Wikipedia pages of notable scientists and their citation-based impact ranking (Samoilenko & Yasseri, 2014). The importance of the editor, as an individual, rather than 'crowd-sourcing' has been discussed, with both subjective interest and self-publicity being behind the motivations of Wikipedia editors (Thelwall, 2018).

Wikipedia has been studied for its network properties, and how this drives growth and quality (Wilkinson & Huberman, 2007). The English language Wikipedia has been identified as being central to the representation of knowledge in multilingual Wikipedia (Ronen et al., 2014). Interestingly, a comparative study of famous people in the Polish and English Wikipedia found systematic differences, relating to the differing cultures, history and values (Callahan & Herring, 2011). There have been criticisms of the coverage of research-based subjects in the non-English language versions of Wikipedia. The number of times a mathematical formula is shared across languages was not found to be a good indicator of quality (Halbach, 2020). For example, articles in ophthalmology were found to range from 'fair' to 'poor' for completeness and organisation, with none being judged 'excellent' (Aguilar-Morales et al., 2021).

Altmetric (part of Digital Science) started collecting citations from the English language Wikipedia in 2015. As of July 2023, Altmetric was indexing 34 languages, including English. In order to index only those languages that have a reasonably diverse and active community of editors, Altmetric applies a threshold of 1000 active editors, as reported by Wikipedia for that language (Wikipedia, 2022b). As Wikipedia languages reach over 1000 active users, they will be added to Altmetric's index. This requirement also supports the exclusion of linguistically inaccurate content, for example the Scots Wikipedia, increasing the likelihood that inaccurate or subject content will be corrected. Furthermore, this number was chosen as a reasonable indicator that novel content existing only in that language would be found, thus avoiding otherwise large Wikipedia that might be created largely by automatic translation.

The Altmetric Manifesto considered Wikipedia to be a possible source of information and data, but was not identified as a potential contributor to the altmetrics (Priem et al, 2010). Nevertheless, in 2013, one-third of bibliometricians felt that Wikipedia was likely to be a good source of impact data (Haustein et al., 2014). By 2015, 5% of PLOS ONE articles had been cited by Wikipedia (Priem et al., 2011). Wikipedia citations are amongst the slowest forms of altmetric data to accumulate (Fang & Costas, 2020). They rarely cite preprints (Fraser et al., 2020), but do cite books at a significant rate (Taylor, 2020). Articles in respected Open Access journals are preferred, and no late growth in preprints has been observed (Benjakob et al., 2022).

The centrality of English has been confirmed by citation studies that compared the Japanese, Chinese and English language Wikipedias (Kikkawa et al., 2016). A study of Spanish and UK articles published in 2012 found that altmetric data sources, including Wikipedia, were insufficient to address language biases in citation data (Mas-Bleda & Thelwall, 2016). Current studies have not examined how the different Wikipedia languages represent scientific discourse through citation, the extent to which they make a unique contribution towards communicating research, and the degree of relationship between them and other Wikipedia. This study aims to fill this gap, with the following research questions:

**Primary objective:** To report the scale of research citation across twenty non-English languages Wikipedia, and the degree to which they cite differing sets of research literature. In particular:

- RQ1. Do the set of papers cited by non-English language Wikipedias differ significantly, both from each other, and from the English language Wikipedia?
- RQ2. To what extent does each non-English language Wikipedia offer a unique representation of research?

**Secondary objective -** To explore and compare subject-area differences across groups, in particular:

- RQ3. Do the set of papers cited by non-English language Wikipedias differ significantly, for each discipline?
- RQ4. To what extent does each non-English language Wikipedia offer a unique representation of research for each discipline?

# Methods

## Design and inclusion criteria

This is an observational cross-sectional study looking at a snapshot of available data from Altmetric and Dimensions (both part of Digital Science) as of 16th August 2022.
For simplicity purposes, we decided to analyse those Wikipedia languages with more than >=10,000 citations. This reduced our dataset from the 31 Wikipedia languages available in Altmetric to twenty non-English Wikipedias, in addition to English.

## Data collection, extraction and database

Data was collected from Altmetric (www.altmetric.com) and Dimensions (www.dimensions.ai) datasets, both part of Digital Science (www.digital-science.com). For data processing, we have used Google BigQuery (GBQ) and the respective Structured Query Language (SQL). Our dataset was defined by identifying all citations from the top 20 Wikipedia languages currently indexed by Altmetric, extracting the language from the domain identifier to create a language-centric model of Wikipedia citations to scholarly

publications – these consisting of journal articles, books and chapters. Using internal identifiers, this dataset was linked to the Dimensions publications dataset.

The **main dataset** includes the following data:

1. Publication IDs (Altmetric, Dimensions, DOI)
2. Publication year
3. Discipline
4. Wikipedia language, and citations mapped to Publication IDs

The Dimensions ingestion process automatically assigns multiple taxonomies to research publications using a machine learning approach which matches documents against taxonomy classes when sufficient confidence is reached [Herzog et al, 2018]. The Fields of Research (FoR) taxonomy is a multi-purpose taxonomy consisting at its highest level, 22 subjects. Where assigned, papers were assigned into six discipline categories (Medical and Health Sciences, Life Sciences, Physical and Mathematical Sciences, Social Sciences, Engineering and Technology and Humanities) based on their computed FoR assignment. Altmetric was founded in 2011 [Bibliometrics: The Leiden Manifesto for research metrics | Nature], and has been indexing numerous data sources for citations, links and mentions, including Wikipedia (11.5M citations). We created a unified dataset from Dimensions and Altmetric to answer our primary objective, which required a matrix comparison between Wikipedia languages and shared publications. An example analysis would be "the number of publications only cited by Spanish Wikipedia *versus* the number of publications cited by both Spanish and English Wikipedias".

## Data and statistical analysis

Each Wikipedia language analysis examines the relationship between three groups:

1. **Group A Publications uniquely cited by the Wikipedia in focus:** this includes the number of publications only cited by a unique non-English language Wikipedia (Region A in Figure 1). Example: publications only cited by Spanish Wikipedia

2. **Group B Publications cited by English Wikipedia:** this includes all the publications cited by a particular language but also shared in the English Wikipedia, Group B in Figure 1. Example: a publication cited both Spanish and English Wikipedias.

   This group also includes publications shared in multiple Wikipedia languages, as long as it includes English Wikipedia, Region B2 in Figure 1. Example: a publication cited by Spanish, English, German and Japanese Wikipedias.

3. **Group C Publications cited only by non-English Wikipedias:** this includes all other publications, included in that particular language, and cited by other Wikipedia languages, *other* than English (Region C in Figure 1). Example: a publication cited by Spanish and Japanese Wikipedias, or Spanish, Japanese and German, but *not* cited by English.

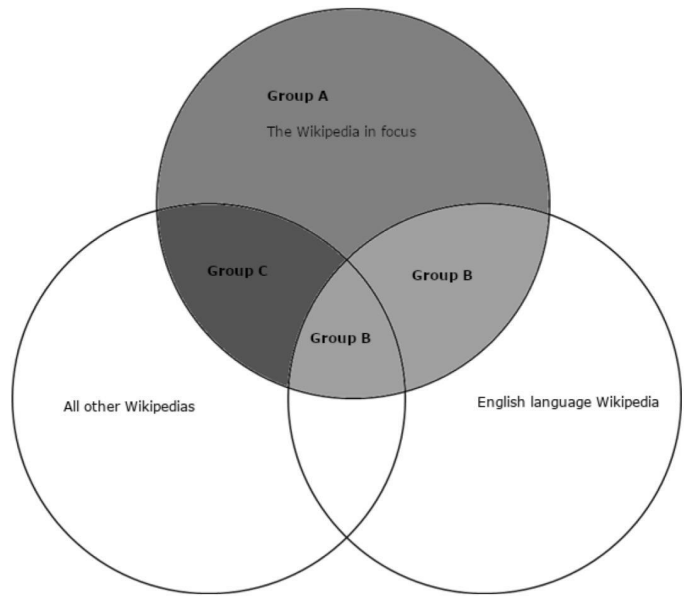

*Figure 1 Schematic illustration of the analysed groups*

We then performed a parametric t-test with Bonferroni correction to compare the three groups, using mean number of publications and percentages, reporting significance ($p < 0.05$). Statistical analysis was performed using R, RStudio and Tidyverse library and were performed individually for each non-English language Wikipedia.

We compared the twenty languages by their proportion of publications cited in Groups A, B and C to understand the extent to which they differ from each other, as potential evidence for exogenous influences on their representation of research. In this analysis, the proportions of each group in each language are compared against the proportion of the global population (i.e. of all twenty languages combined) and analysed using Excel's Chi-square function.

### Sub-analysis

Sub-analysis of the primary and secondary outcomes were applied to the different cohorts, defined by subject area of the cited literature. Similar descriptive and statistical analysis methods were conducted.

# Results

Of the 31 languages indexed by Altmetric, 20 languages and English were analysed, citing a total of 3,485,474 distinct publications. The number of distinct publications per Wikipedia language can be found in Table 1 and the degree of intersection between Wikipedia languages in Table 2.

*Table 1 Age, size and number of cited research publications of Wikipedia by language*

| Language | Launch date | Number of Wikipedia pages | Ranking by number of Wikipedia pages | Cited publications | Ranking by number of cited publications |
|---|---|---|---|---|---|
| | | | | | |

| Language | Date | Articles | Rank | Unique research publications | Rank |
|---|---|---|---|---|---|
| Arabic | July 2003 | 1,212,000 | 11 | 302,618 | 5 |
| Catalan | March 2001 | 681,000 | 15 | 188,414 | 11 |
| Czech | May 2002 | 484,000 | 19 | 85,724 | 20 |
| English | January 2001 | 6,635,834 | 1 | 2,035,466 | 1 |
| Farzi (or Persian) | December 2003 | 818,000 | 14 | 113,431 | 17 |
| French | March 2001 | 2,375,000 | 3 | 421,388 | 2 |
| German | March 2001 | 2,618,000 | 2 | 279,238 | 6 |
| Greek | December 2002 | 199,000 | 21 | 84,382 | 21 |
| Indonesian | May 2003 | 583,000 | 17 | 157,058 | 14 |
| Italian | May 2001 | 1,738,000 | 5 | 278,222 | 7 |
| Japanese | May 2001 | 1,342,000 | 8 | 263,044 | 8 |
| Korean | October 2002 | 574,000 | 18 | 104,822 | 18 |
| Mandarin | October 2002 | 1,221,000 | 10 | 251,289 | 9 |
| Polish | September 2001 | 1,507,000 | 7 | 135,604 | 16 |
| Portuguese | May 2001 | 1,098,000 | 13 | 233,211 | 10 |
| Russian | May 2001 | 1,787,000 | 4 | 309,832 | 4 |
| Serbian | February 2003 | 659,000 | 16 | 159,080 | 13 |
| Spanish | November 2001 | 1,711,000 | 6 | 373,564 | 3 |
| Turkish | December 2002 | 413,000 | 20 | 94,002 | 19 |
| Ukrainian | January 2004 | 1,112,000 | 12 | 146,411 | 15 |
| Vietnamese | November 2002 | 1,328,000 | 9 | 164,527 | 12 |

*Table 2 Intersection between papers cited by Wikipedia languages, expressed as an absolute number of unique research publications (above the line), and as a percentage of the total unique research publications (below the white cells)*

|            | Arabic | Catalan | Czech | English | French | German | Greek | Indonesian | Italian | Japanese | Korean | Mandarin | Persian | Polish | Portuguese | Russian | Serbian | Spanish | Turkish | Ukrainian | Vietnamese |
|------------|--------|---------|-------|---------|--------|--------|-------|------------|---------|----------|--------|----------|---------|--------|------------|---------|---------|---------|---------|-----------|------------|
| Arabic     |        | 42696   | 19134 | 226725  | 57017  | 28126  | 27630 | 46674      | 52385   | 43113    | 28008  | 61556    | 36308   | 21027  | 57531      | 53501   | 40948   | 73165   | 30735   | 35018     | 50329      |
| Catalan    | 1.20%  |         | 14650 | 97488   | 49555  | 21189  | 18247 | 33569      | 39200   | 29465    | 20527  | 40063    | 20118   | 17271  | 41568      | 39595   | 30086   | 70364   | 20532   | 22922     | 32680      |
| Czech      | 0.50%  | 0.40%   |       | 47231   | 21370  | 12460  | 9560  | 15606      | 18319   | 14812    | 9602   | 20053    | 10607   | 10392  | 17758      | 18256   | 14005   | 22045   | 9940    | 12011     | 15150      |
| English    | 6.50%  | 2.80%   | 1.40% |         | 167578 | 89106  | 57793 | 102510     | 127885  | 128524   | 65349  | 155110   | 81825   | 50542  | 147071     | 146739  | 103795  | 207711  | 64629   | 89505     | 116653     |
| French     | 1.60%  | 1.40%   | 0.60% | 4.80%   |        | 42840  | 22108 | 42783      | 57763   | 44333    | 25107  | 54941    | 25394   | 24916  | 59112      | 58211   | 35122   | 84208   | 25572   | 31168     | 40667      |
| German     | 0.80%  | 0.60%   | 0.40% | 2.60%   | 1.20%  |        | 11109 | 19074      | 28073   | 23694    | 12155  | 28293    | 13268   | 14966  | 26079      | 29913   | 16611   | 36194   | 13047   | 16625     | 19722      |
| Greek      | 0.80%  | 0.50%   | 0.30% | 1.70%   | 0.60%  | 0.30%  |       | 21729      | 21728   | 16626    | 12549  | 24773    | 14393   | 9644   | 23584      | 21735   | 19545   | 26451   | 14668   | 14499     | 21869      |
| Indonesian | 1.30%  | 1.00%   | 0.40% | 2.90%   | 1.20%  | 0.50%  | 0.60% |            | 38429   | 30488    | 22287  | 46439    | 25269   | 17075  | 43946      | 40430   | 36574   | 50758   | 23445   | 23785     | 40754      |
| Italian    | 1.50%  | 1.10%   | 0.50% | 3.70%   | 1.70%  | 0.80%  | 0.60% | 1.10%      |         | 37027    | 22720  | 48927    | 23833   | 21022  | 47516      | 47975   | 31596   | 63316   | 23837   | 27542     | 37444      |
| Japanese   | 1.20%  | 0.80%   | 0.40% | 3.70%   | 1.30%  | 0.70%  | 0.50% | 0.90%      | 1.10%   |          | 23721  | 48079    | 21202   | 16390  | 39957      | 41734   | 28816   | 51823   | 18578   | 23517     | 33093      |
| Korean     | 0.80%  | 0.60%   | 0.30% | 1.90%   | 0.70%  | 0.30%  | 0.40% | 0.60%      | 0.70%   | 0.70%    |        | 28794    | 14339   | 10217  | 26176      | 24474   | 19292   | 28676   | 14110   | 14914     | 24279      |
| Mandarin   | 1.80%  | 1.10%   | 0.60% | 4.50%   | 1.60%  | 0.80%  | 0.70% | 1.30%      | 1.40%   | 1.40%    | 0.80%  |          | 29399   | 21452  | 54826      | 51688   | 39387   | 65429   | 26966   | 30713     | 49149      |
| Persian    | 1.00%  | 0.60%   | 0.30% | 2.30%   | 0.70%  | 0.40%  | 0.40% | 0.70%      | 0.70%   | 0.60%    | 0.40%  | 0.80%    |         | 10025  | 27191      | 25759   | 23123   | 30889   | 16491   | 17428     | 24272      |
| Polish     | 0.60%  | 0.50%   | 0.30% | 1.50%   | 0.70%  | 0.40%  | 0.30% | 0.50%      | 0.60%   | 0.50%    | 0.30%  | 0.60%    | 0.30%   |        | 21162      | 23795   | 14182   | 26573   | 10929   | 15264     | 16984      |
| Portuguese | 1.70%  | 1.20%   | 0.50% | 4.20%   | 1.70%  | 0.70%  | 0.70% | 1.30%      | 1.40%   | 1.10%    | 0.80%  | 1.60%    | 0.80%   | 0.60%  |            | 50118   | 37118   | 69394   | 25898   | 28984     | 44168      |
| Russian    | 1.50%  | 1.10%   | 0.50% | 4.20%   | 1.70%  | 0.90%  | 0.60% | 1.20%      | 1.40%   | 1.20%    | 0.70%  | 1.50%    | 0.70%   | 0.70%  | 1.40%      |         | 36624   | 64604   | 24245   | 51989     | 39523      |
| Serbian    | 1.20%  | 0.90%   | 0.40% | 3.00%   | 1.00%  | 0.50%  | 0.60% | 1.00%      | 0.90%   | 0.80%    | 0.60%  | 1.10%    | 0.70%   | 0.40%  | 1.10%      | 1.10%   |         | 44309   | 20553   | 22175     | 36436      |
| Spanish    | 2.10%  | 2.00%   | 0.60% | 6.00%   | 2.40%  | 1.00%  | 0.80% | 1.50%      | 1.80%   | 1.50%    | 0.80%  | 1.90%    | 0.90%   | 0.80%  | 2.00%      | 1.90%   | 1.30%   |         | 30541   | 36295     | 49239      |
| Turkish    | 0.90%  | 0.60%   | 0.30% | 1.90%   | 0.70%  | 0.40%  | 0.40% | 0.70%      | 0.70%   | 0.50%    | 0.40%  | 0.80%    | 0.50%   | 0.30%  | 0.70%      | 0.70%   | 0.60%   | 0.90%   |         | 15804     | 23833      |
| Ukrainian  | 1.00%  | 0.70%   | 0.30% | 2.60%   | 0.90%  | 0.50%  | 0.40% | 0.70%      | 0.80%   | 0.70%    | 0.40%  | 0.90%    | 0.50%   | 0.40%  | 0.80%      | 1.50%   | 0.60%   | 1.00%   | 0.50%   |           | 23802      |
| Vietnamese | 1.40%  | 0.90%   | 0.40% | 3.30%   | 1.20%  | 0.60%  | 0.60% | 1.20%      | 1.10%   | 0.90%    | 0.70%  | 1.40%    | 0.70%   | 0.50%  | 1.30%      | 1.10%   | 1.00%   | 1.40%   | 0.70%   | 0.70%     |            |

## English Wikipedia

Of all the languages analysed, English is the dominant 2,035,466 (58.4%) of the 3,485,474 total publications (Table 1). In terms of absolute numbers, Spanish and Arabic have the largest overlap with English (Table 2), having 227k and 207k in common (respectively 6.5% and 6.0% of the complete corpus of cited literature). Of all publications cited by the English language Wikipedia, 1140160 (32.7%) are uncited by other languages analysed in this study. Non-English language Wikipedias cite an additional 1,457,138 (or 41.8%) papers, these not being by English Wikipedia. The size of Wikipedia by volume of pages correlates strongly with the number of cited publications (R = 0.95).

## Non-English Wikipedias

A total of 20 non-English Wikipedia languages were included in the analysis. The total number and percentage of publications cited for each group in all included Wikipedia languages can be found in Figures 2 and 3. Descriptive statistics for Unique, English and Other languages groups can be found in Table 3.

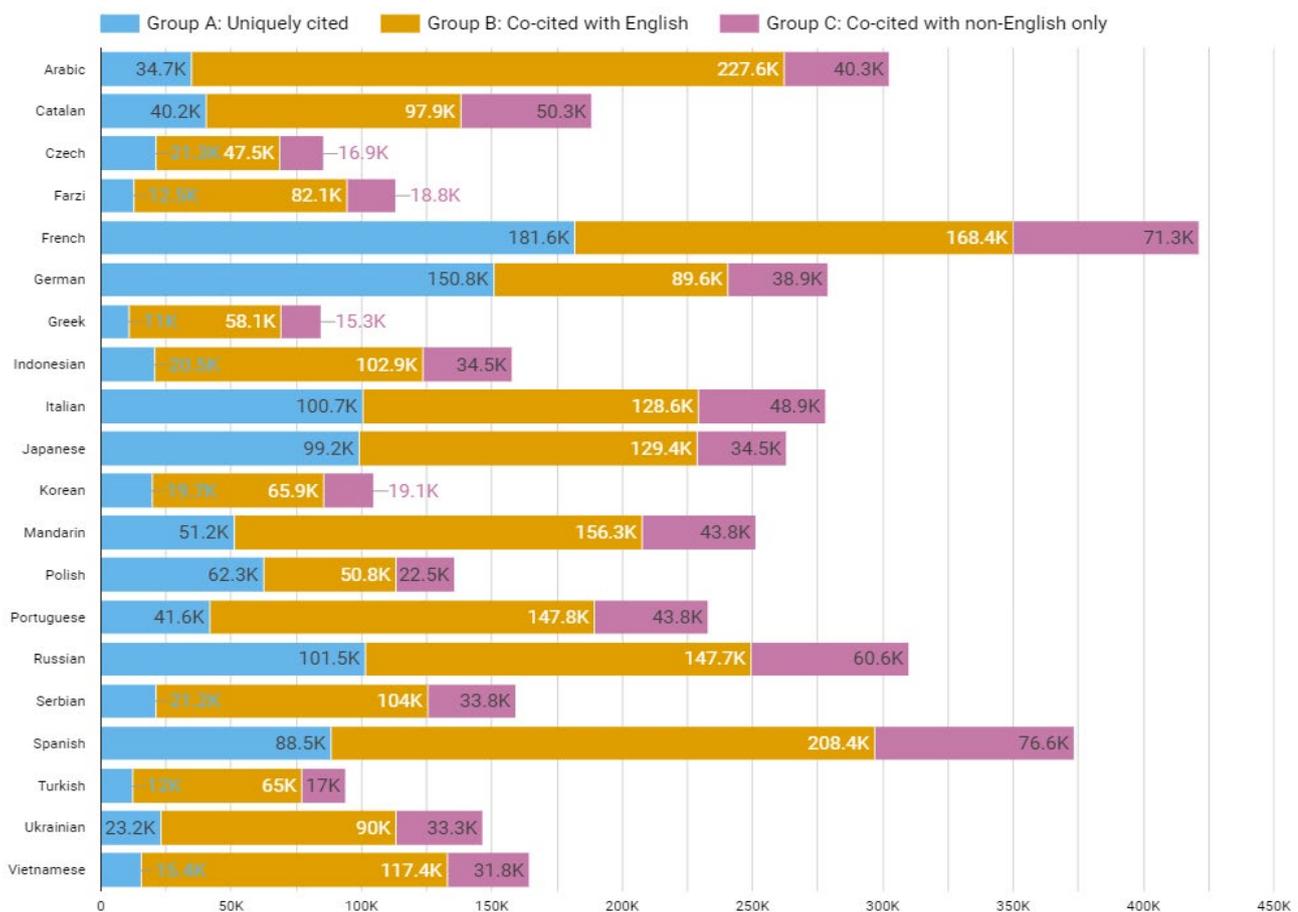

Figure 2 Total number of publications cited in each group, by language.

French, Spanish, Russian, Arabic, German and Italian have the highest number of cited publications (Figure 3a). When compared to the size of the overall Wikipedias (Table 1), only Arabic appears to have a higher-than-expected density of citations – all other languages are similarly placed in both the top-third of citing Wikipedia and the top-third of Wikipedia size.

Of the non-Latin alphabet Wikipedias, Chinese, Arabic and Russian cite the largest number of publications. The smallest Wikipedia, in terms of cited research corpus, are Czech, Farzi, Greek, Korean and Turkish. This corresponds with their position in the bottom third when calculated by overall Wikipedia size.

Spanish and Arabic both share over 200,000 citations with English (Group B). Group B is the largest group within each language, except for German, French and Polish, where Group A dominates. French, German, Russian and Italian all uniquely cite over 100,000 publications (Group A), with Japanese being just under 100,000 unique citations. A number of Wikipedias cite a very small number of unique papers (Farzi, Greek, Turkish and Vietnamese cite <= 15,000)

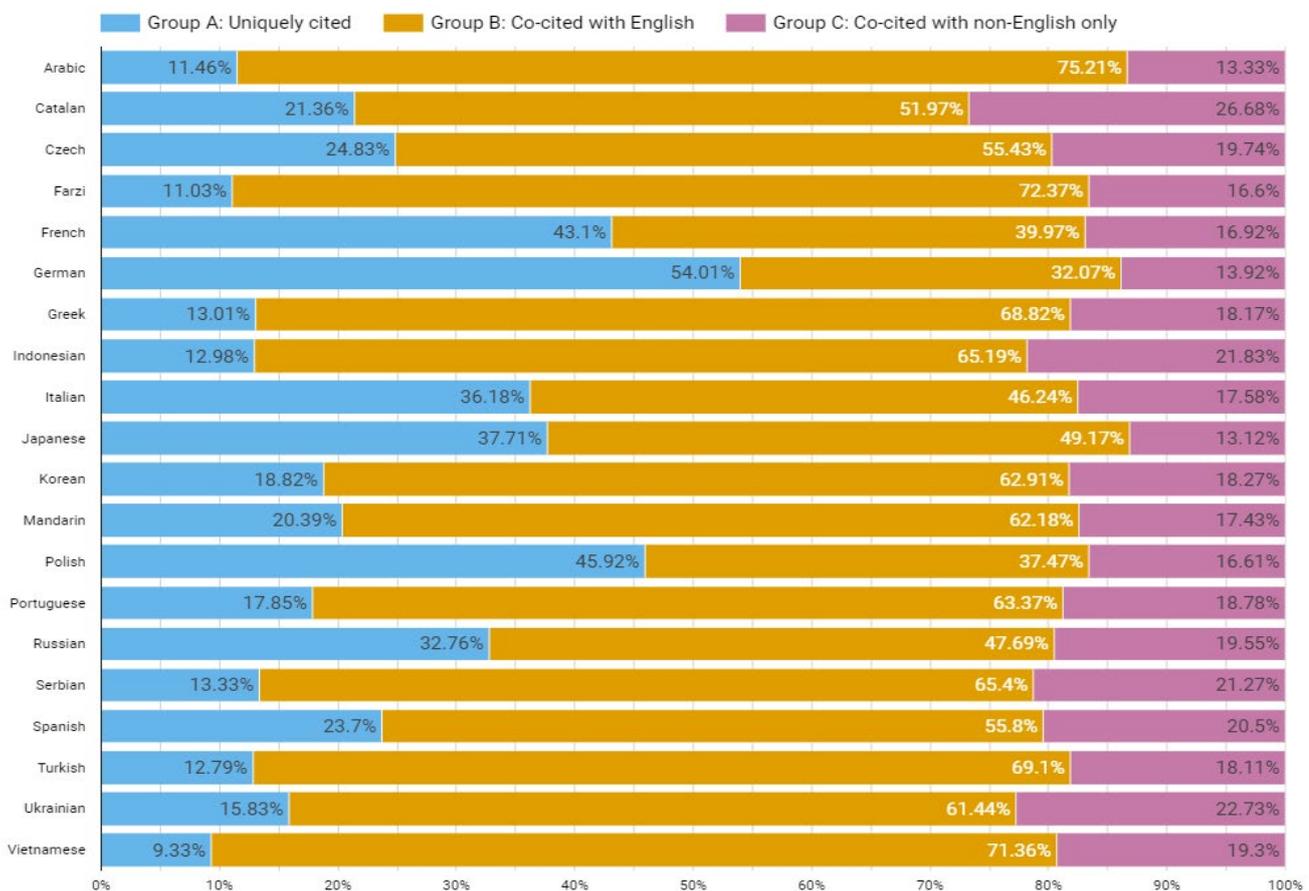

Figure 3 Proportion of publications cited in each group, by language.

The different Wikipedias show a wide variety of proportions between Groups A, B and C, with the lowest variations exhibited for Group C: Catalan (27%) shows the highest proportion of co-citation with other non-English Wikipedias, and Arabic the lowest, with 13% (Figure 3b). In contrast, the variation in Group A (uniquely cited research) is higher, with Vietnamese showing a proportion < 10%, and German being 54%. French and Polish have the lowest proportion of published research shared with English (37% and 40%), with Arabic, Farzi and Vietnamese showing the highest proportion, with 75%, 72% and 71% in common with English (Group B).

In general, the smaller Wikipedias show the highest levels of co-citation with English (Group B), and the lowest levels of unique citations (Group A). The larger Wikipedias show a great variation: whereas French is larger than German in terms of research cited, German shows a higher proportion of uniquely cited research. However, Polish – one of the smaller Wikipedias, in terms of volume of citations – is one of the most distinctive, with the second highest proportion of uniquely cited research. In contrast, Spanish is second largest Wikipedia, but uniquely cites a relatively small proportion of unique literature.

There was a high statistically significant difference between the number of publications cited only by a unique Wikipedia (Group A), the number of publications cited by the English Wikipedia (Group B), ($p <0.001$), and the number of publications cited by English Wikipedia (Group B) and other (non-English) Wikipedias ($p<0.001$), (Group C) (Table 3). There was no statistical significance between the number of publications only shared in a unique Wikipedia (Group A) and other (non-English) Wikipedias ($p>0.05$) (Group C). The same is applicable to the respective percentages.

*Table 3 - Wikipedia cited publications descriptive statistics*

| Metric/Group | Group A - Unique | Group B - English | Group C - Other |
| --- | --- | --- | --- |
| Mean ± SD publications cited in Wikipedia | 54433 ± 50332 | 114273 ± 50390 | 38586 ± 18000 |
| Percentage publications cited in Wikipedia | 23.2% ± 14.1% | 57.7% ± 12.4% | 19.2% ± 4.8% |

Table 4 presents the proportions of Groups A, B and C by language, and a Chi-squared value that compares an individual's proportions against the overall group percentage. Several languages show significant deviations from the global proportion. Arabic, Farzi and Vietnamese both have a higher than expected proportion of papers in common with the English Wikipedia and lower than expected for non-English Wikipedia, and thus significantly deviate from the expected distribution when analysed by a Chi-squared test for population. French, German and Polish - which have higher levels of uniquely cited publications than expected, and less in common with non-English Wikipedia. These three languages also deviate significantly from the global population.

*Table 4 Proportion of Groups A, B, C for each language, and for the global population of all twenty Wikipedias, with Chi-Squared values for population*

|  | Group A Unique | Group B Cited with English | Group C Cited with non-English | Chi-squared |
| --- | --- | --- | --- | --- |
| All 20 languages | 26% | 55% | 19% | n/a |
| Arabic | 11% | 75% | 13% | 0.92 |
| Catalan | 21% | 52% | 27% | 0.98 |

| Language | | | | |
|---|---|---|---|---|
| Czech | 25% | 55% | 20% | 1.00 |
| Farzi | 11% | 72% | 17% | 0.93 |
| French | 43% | 40% | 17% | 0.93 |
| German | 54% | 32% | 14% | 0.82 |
| Greek | 13% | 69% | 18% | 0.95 |
| Indonesian | 13% | 65% | 22% | 0.96 |
| Italian | 36% | 46% | 18% | 0.97 |
| Japanese | 38% | 49% | 13% | 0.96 |
| Korean | 19% | 63% | 18% | 0.98 |
| Mandarin | 20% | 62% | 17% | 0.99 |
| Polish | 46% | 37% | 17% | 0.90 |
| Portuguese | 18% | 63% | 19% | 0.98 |
| Russian | 33% | 48% | 20% | 0.99 |
| Serbian | 13% | 65% | 21% | 0.96 |
| Spanish | 24% | 56% | 21% | 1.00 |
| Turkish | 13% | 69% | 18% | 0.95 |
| Ukrainian | 16% | 61% | 23% | 0.97 |
| Vietnamese | 9% | 71% | 19% | 0.92 |

Analysis by Discipline

We have subdivided by discipline, assigning each publication into six categories based on their assigned Fields of Research code, being Engineering and Technology (E&T), Humanities (H), Life Sciences (LS), Medical and Health Sciences (MHS), Physics and Maths (P&M), and Social Sciences (SS).

Many languages show little variation of breakdown (Figure 4). Although many of these are the smaller Wikipedias (e.g. Farzi, Serbian, Vietnamese), this characteristic is also shared by larger Wikipedia (e.g. Mandarin, Arabic). Catalan has a notably high proportion of uniquely-cited MHS research, as does German, Polish, Italian and French Wikipedias. Ukranian has a high proportion of uniquely-cited LS research, and a high proportion of E&T shared with non-English language Wikipedias. Only Spanish shows a relatively high proportion of uniquely cited H research; although uniquely cited SS publications are disproportionately cited by Czech, French, Polish and French Wikipedia.

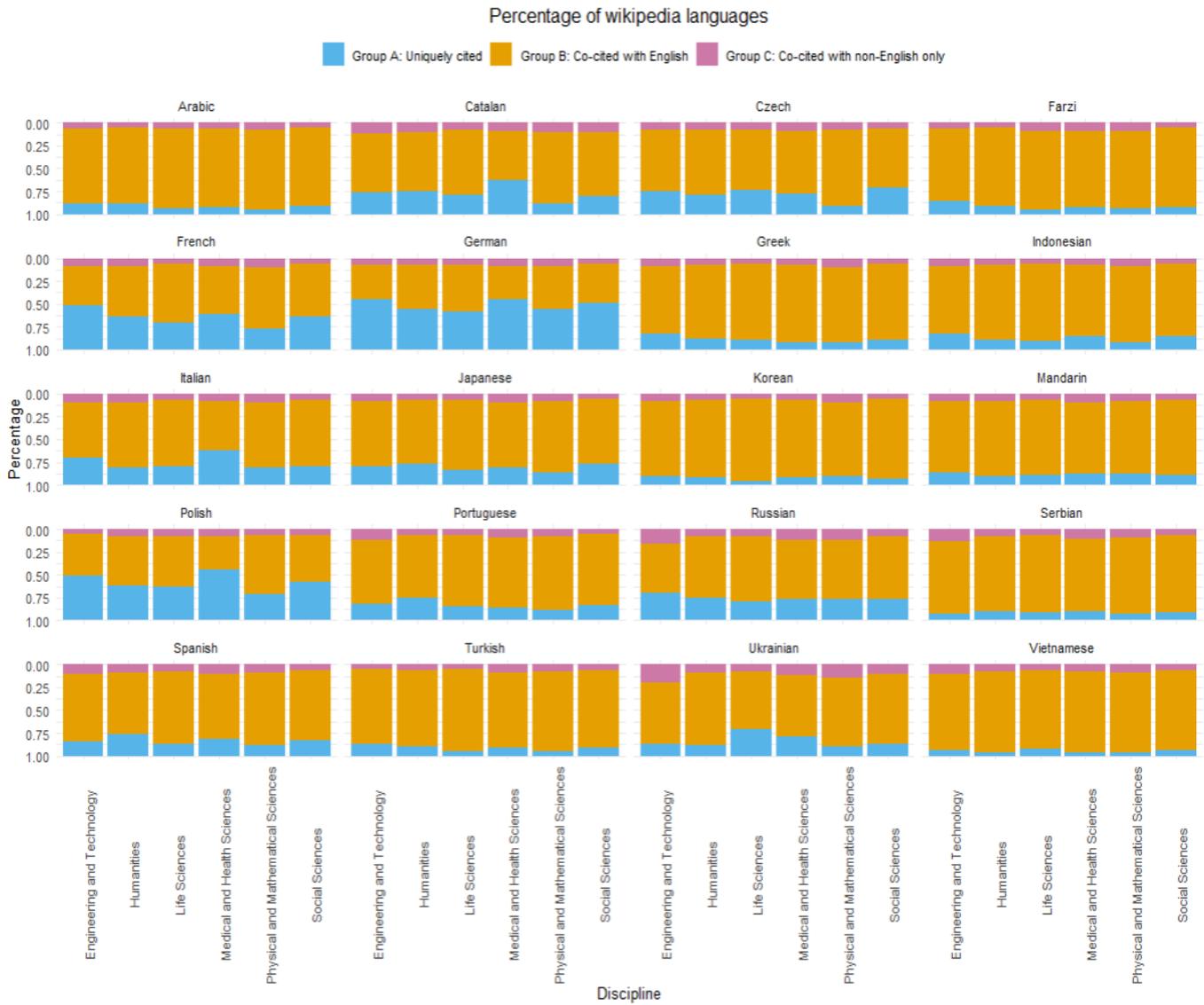

Figure 4 - Percentages of publications included in each study group, by discipline and country.

When comparing all three groups inside each discipline, statistical significant differences can be found amongst almost all groups (table 5).

Table 5 - Statistical comparison table between groups across disciplines. NS - not significant, p>0.05, * p<0.05, ** p<0.01, *** p<0.001, **** p<0.0001

| Discipline | English/Other (Group B vs Group C) | English/Unique (Group B vs Group A) | Other/Unique (Group A vs Group C) |
|---|---|---|---|
| Engineering and Technology | **** | **** | ** |
| Humanities | **** | **** | ** |
| Life Sciences | **** | **** | ** |
| Medical | **** | **** | * |
| Physics and Maths | **** | **** | NS |
| Social Sciences | **** | **** | ** |

We also compared all disciplines for each of the study groups, with most disciplines not being statistically significant inside the percentage of outputs shared in English Wikipedia, as well as in Unique Wikipedias. There were some statistically significant differences between some disciplines in the percentage of outputs in other (non-English) Wikipedias (Table 6). For example the amount of publications cited in non-English wikipedias is significantly different between Engineering and Technology and Humanitites (p=0.039), in Table 6.

Table 6 - Statistical comparison across disciplines in each study group. NS - not significant, p>0.05, * p<0.05, ** p<0.01.

| *Discipline* | Engineering and Technology | Humanities | Life Sciences | Medical | Physics and Maths | Social Sciences |
|---|---|---|---|---|---|---|
| Engineering and Technology | NA | English (B): 0.231 **Other (C): 0.039*** Unique (A): 0.523 | English: 0.071 **Other: 0.005*** Unique: 0.296 | English: 0.757 Other: 0.491 Unique: 0.899 | **English: 0.038*** Other: 0.513 Unique: 0.066 | English: 0.148 **Other: 0.003*** Unique: 0.510 |
| Humanities | | NA | English: 0.537 Other: 0.452 Unique: 0.683 | English: 0.374 Other: 0.166 Unique: 0.609 | English: 0.375 Other: 0.156 Unique: 0.228 | English: 0.800 Other: 0.365 Unique: 0.985 |
| Life Sciences | | | NA | English: 0.133 **Other: 0.034*** Unique: 0.358 | English: 0.786 **Other: 0.031*** Unique: 0.423 | English: 0.716 Other: 0.877 Unique: 0.697 |
| Medical | | | | | NA | English: 0.077<br>Other: 0.365<br>Unique: 0.087 | English: 0.254<br>**Other: 0.023*** <br>Unique: 0.595 |
| Physics and Maths | | | | | | NA | English: 0.525<br>**Other: 0.021*** <br>Unique: 0.523 |
| Social Sciences | | | | | | | NA |

# Discussion

This study aimed to report the scale of research citation across twenty non-English languages Wikipedia, and the degree to which they cite a different set of research literature across six different disciplines. Our results suggest a significant difference between both the number and percentage of publications shared in English (group B) vs non-English Wikipedias (both in a unique (group A) language, or multiple non-English (group C)). This finding – that while English has greater coverage, other languages show significant uniqueness-of-voice - is in contrast to previously reported research (Ronen et al., 2014). Our results also indicate that approximately 40% of publications are only cited in non-English language Wikipedias, either in a single Wikipedia language (≈23.2%) or in more than one non-English Wikipedia (≈19.2%), slightly varying by country and discipline. This answers our primary research question, with a strong indication that the aforementioned uniqueness-of-voice of non-English Wikipedias is present.

Smaller non-English Wikipedias may be considered to show a less unique view of science than their larger counterparts. That noted, there are outliers: Polish, for example, is smaller than Ukrainian but shows considerably more uniqueness-of-voice, suggesting that there may also be a relationship with other factors, for example, editorial direction as previously identified (Samoilenko & Yasseri, 2014), interest of individual contributors (Thelwall, 2018) or maturity of Wikipedia. Non-Latin character sets may not be considered to be either a factor in, or an obstacle to, citation analysis, with Russian and Japanese both showing high rates of citation and significant uniqueness-of-voice, as analysed by citations. When considering non-English language Wikipedias, we cannot forget the importance and meaning of language, identity and the role that Wikipedia has in communicating information to a population. Both the Catalan and Ukrainian languages may be considered as being tightly associated with Catalan identity and political voice, and with the survival and identity of the country of Ukraine. Population size and linguistic diversity may also be important: Indonesian is spoken by 23 million people, but has a relatively small Wikipedia.

We have also compared the percentages for each of six disciplines across the twenty Wikipedias and have found a degree of consistency across the languages. Across the Wikipedias that show great diversity from English, we generally find that both Medical and Health Sciences, and Engineering and Technology show a greater degree of uniqueness of voice than other fields. Within other languages, we see individual differences. Czech and Japanese have more diversity in the Social Sciences, and Ukrainian in Life Sciences. Thus, even in the case of a non-English language

Wikipedia with low rates of diversity from English, significant differences may be observed within disciplines. These differences may be explained by the focus of contributors, the expertise of editors or through the particular importance of these disciplines for countries writing in these languages. This high degree of diversity in these two fields has not been previously reported, although the centrality of English to the Humanities has been observed (Ronen et al., 2014).

Our research suggests that as representations of scientific discourse vary by language, so may the representation of a subject vary, with direct comparison between individual pages proving challenging. For example, the representation of a single person – in this case, the eighteenth century French scholar, Emilie du Chatelet – is marked with 41 scholarly citations in the English language Wikipedia, whereas the French equivalent page has only 30 scholarly citations (The French language Wikipedia page also contains many more links to popular cultural representations and webpages). However the French language Wikipedia has dedicated pages to du Chatalet's works, for example on *Discours sur le bonheur* – the page of which has 63 scholarly citations. The English language Wikipedia has no such equivalent page.

There have been proposals to translate the English language Wikipedia into other languages, however, knowledge representation across languages is more than simple like-for-like translation, the act of translation may not be treated simply as the codification of facts (Colina et al., 2022), and thus the complexities of translation cannot be underestimated. A further complexity can be seen in the example discussed above: can any language be considered canonical, especially when the 'stem' article (for example, on du Chatelet) may be longer in the English Wikipedia, but the overall representation of du Chatelet be richer in French, with multiple pages dedicated to discussing her work, life, and meaning within Francophone cultures. Indeed, widescale translation from English to non-English Wikipedia may be seen as an 'othering' of both other languages, other cultures and other representations of research-linked knowledge.

## Limitations

One limitation of the analysis of this study is that, for readability and simplicity of this manuscript, we reduced the number of included languages to 21, with a criteria of a minimum of 10,000 citations to be included. This design decision may have hidden some data that might have been valuable to the study, however we believe that the low number of citations would not have made a significant change to our results and conclusion. Secondly, this study does not distinguish between citations to books, chapters, and journal articles. A separate paper will analyse these separately.

## Further Work

This research suggests further analysis of the Wikipedia corpus:

- The extent to which Open Access publications and preprints are involved in the differences reported here, OA publication rates vary by both region and subject.
- The relationship between modes of publication and Wikipedia citations, in particular the relative importance of book versus journal content.
- Possible exogenous factors that might drive citation behaviour whether through cultural struggle, political or economic change or contribution of individual editors.

- The citing pages of this study are not analysed by this study. Wikipedia citations may be made by a variety of types of page: for example, biographical pages, topic reviews and specific subject pages all make use of academic citations.

# Conclusion

This study included almost 4 million distinct publications, with over 1 million journal papers and more than 2 million books. To our knowledge, this is the biggest study ever conducted in Wikipedia citations. The results are clear: although English cites the largest number of publications (2M) by a considerable margin, giving the most complete view of scholarly outputs, as measured by citations, its total of uniquely cited research (1.1M publications) is smaller than the total of uniquely cited research (1.4M) covered by the other 20 Wikipedias studied here. Thus we can confidently report that these other Wikipedias may provide a unique model of scientific discourse: non-English Wikipedias cannot generally be treated as either translations or subsets of English Wikipedia. There is a variety of unique research being referenced in non-English Wikipedias.